\documentclass[twocolumn,showpacs,amsmath,amssymb,prb]{revtex4}
\usepackage{graphicx}
\usepackage{bm}
\usepackage{natbib}

\begin{document}
\title{Aharonov-Bohm oscillations in coupled quantum dots: Effect of electron-electron interactions}
\author{Andrew G. Semenov}
 \email{semenov@lpi.ru}
\affiliation{I.E. Tamm Department of Theoretical Physics, P.N.
Lebedev Physics Institute, 119991 Moscow, Russia}
\author{Dmitri S. Golubev}
\affiliation{Forschungszentrum Karlsruhe, Institut f\"ur
Nanotechnologie, 76021, Karlsruhe, Germany}
\author{Andrei D. Zaikin}
\affiliation{Forschungszentrum Karlsruhe, Institut f\"ur
Nanotechnologie, 76021, Karlsruhe, Germany} \affiliation{I.E. Tamm
Department of Theoretical Physics, P.N. Lebedev Physics Institute,
119991 Moscow, Russia}

\begin{abstract}
We theoretically analyze the effect of electron-electron
interactions on Aharonov-Bohm (AB) current oscillations in
ring-shaped systems with metallic quantum dots pierced by
external magnetic field.  We demonstrate that electron-electron
interactions suppress the amplitude of AB oscillations $I_{AB}$ at
all temperatures down to $T=0$ and formulate quantitative
predictions which can be verified in future experiments. We argue
that the main physical reason for such interaction-induced
suppression of $I_{AB}$ is electron dephasing while Coulomb
blockade effects remain insignificant in the case of metallic
quantum dots considered here. We also emphasize a direct
relation between our results and the so-called $P(E)$-theory
describing tunneling of interacting electrons.

\end{abstract}


\maketitle
\section{Introduction}
Aharonov-Bohm (AB) oscillations of conductance as a function of
the magnetic flux $\Phi$ piercing the system represent one of the
fundamental properties of meso- and nanoscale conductors which is
directly related to {\it quantum coherence} of electrons
\cite{ArSh}. Coherent electrons propagating along different paths
in multiply connected conductors, such as, e.g., metallic rings,
can interfere. Such interference effect results in a specific
quantum contribution to the system conductance $\delta G$.
Threading the ring by an external magnetic flux $\Phi$ one can
control the relative phase of the wave functions of interfering
electrons, thus changing the magnitude of $\delta G$ as a function
of $\Phi$. The dependence $\delta G (\Phi )$ turns out to be
periodic with the fundamental period equal to the flux quantum
$\Phi_0=hc/e$.

It is important to emphasize that the phase of the electron wave
function is sensitive to its particular path. In diffusive
conductors electrons can propagate along very many different
paths, hence picking up different phases. Averaging over these
(random) phases or, equivalently, over disorder configurations
yields the amplitude of AB oscillations $\delta G (\Phi )$ with
the period $\Phi_0$ to vanish in diffusive conductors \cite{ArSh}.
There exists, however, a special class of electron trajectories
which interference is not sensitive to disorder averaging. These
are all pairs of time-reversed paths which are also responsible for the
phenomenon of weak localization \cite{CS}. In multiply connected
disordered conductors interference between these trajectories
gives rise to non-vanishing AB oscillations with the principle
period $\Phi_0/2$. Such oscillations will be analyzed below in
this paper.

It is well known that various kinds of interactions, such as
electron-electron and electron-phonon interactions, electron
scattering on magnetic impurities etc. can lead to decoherence of
electrons thus reducing their ability to interfere. Accordingly,
AB oscillations should be sensitive to all these processes and can
be used as a tool to probe the fundamental effect of interactions
on quantum coherence of electrons in nanoscale conductors.
Recently it was demonstrated \cite{GZ06,GZ08,GZ07} that the effect
of quantum decoherence by electron-electron interactions can be
conveniently studied employing the model of a system of coupled
quantum dots (or scatterers). This model might embrace essentially
all types of disordered conductors and allows for a
straightforward non-perturbative treatment of electron-electron
interactions. It also allows to establish a direct and transparent
relation \cite{GZ08,GZ07} between the problem of quantum
decoherence by electron-electron interactions and the so-called
$P(E)$ theory \cite{SZ,IN}, see also \cite{GZ99} for an earlier
discussion of this important point. In this paper we employ a
similar model in order to study the effect of electron-electron
interactions on AB oscillations in disordered nanorings.

The structure of our paper is as follows. In Sec. 2 we define our
model and outline our general real time path integral formalism
employed in this work. Sec. 3 is devoted to a detailed derivation
of the effective action for our problem in terms of fluctuating
Hubbard-Stratonovich fields mediating electron-electron
interactions. With the aid of this effective action we then
evaluate Aharonov-Bohm conductance of the ring in the presence of
electron-electron interactions. This task is accomplished in Sec.
4. A brief discussion of our results is presented in Sec. 5. Some
technical details of disorder averaging are relegated to Appendices.

\section{The model and basic formalism}

Below we will analyze the system depicted in Fig. 1. The structure
consists of two chaotic quantum dots (L and R) characterized by
mean level spacing $\delta_L$ and $\delta_R$. Here we will
restrict our attention to the case of metallic quantum dots with
$\delta_{L,R}$ being the lowest energy parameters in the problem.
These dots are interconnected via two tunnel junctions J$_1$ and
J$_2$ with conductances $G_{t1}$ and $G_{t2}$ forming a
ring-shaped configuration as shown in Fig. 1. The left and right
dots are also connected to the leads (LL and RL) respectively via
the barriers J$_L$ and J$_R$ with conductances $G_L$ and $G_R$. We
also define the corresponding dimensionless conductances of all
four barriers as $g_{t1,2}=G_{t1,2}R_q$ and $g_{L,R}=G_{t1,2}R_q$,
where $R_q=2\pi /e^2$ is the quantum resistance unit. These
dimensionless conductances are related to the barrier channel
transmissions $T_k$ via the standard formula $g =2\sum_kT_k$,
where the sum is taken over all conducting channels in the
corresponding barrier and an extra factor 2 accounts for the
electron spin.

For the sake of convenience in what follows we will assume that
dimensionless conductances $g_{L,R}$ are much larger than unity,
while the conductances $g_{t1}$ and $g_{t2}$ are small as compared
to those of the outer barriers, i.e.
\begin{equation}
g_L,g_R\gg 1,g_{t1},g_{t2}. \label{met}
\end{equation}
 The whole structure is pierced by the magnetic
flux $\Phi$ through the hole between two central barriers in such
way that electrons passing from left to right through different
junctions acquire different geometric phases. Applying a voltage
across the system one induces the current which shows AB
oscillations with changing the external flux $\Phi$.

\begin{figure}
 \centering
\includegraphics[width=2.5in]{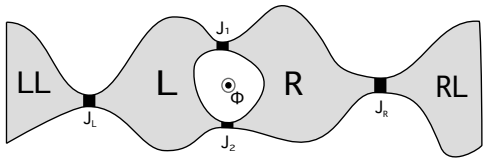}
\caption{\label{f1}The ring-shaped quantum dot structure under
consideration.}
\end{figure}

The system depicted in Fig. 1 is described by the effective
Hamiltonian:
\begin{eqnarray}
\hat H&=&\sum_{i,j=L,R}\frac{C_{ij}\hat {\bm V}_i\hat {\bm
V}_j}{2} +\hat {\bm H}_{LL}+\hat {\bm H}_{RL} \nonumber\\ &&
+\sum_{j=L,R}\hat {\bm H}_{j} +\hat{\bm T}_L +\hat {\bm T}_R+ \hat
{\bm T}, \label{H}
\end{eqnarray}
where $C_{ij}$ is the capacitance matrix, $\hat {\bm V}_{L(R)}$ is
the electric potential operator on the left (right) quantum dot,
$$
\hat {\bm
H}_{LL}=\sum\limits_{\alpha=\uparrow,\downarrow}\int\limits_{LL}d^3{\bm
r} \hat\Psi^\dagger_{\alpha,LL}({\bm r})(\hat
H_{LL}-eV_{LL})\hat\Psi_{\alpha,LL}({\bm r}),
 $$
$$
\hat {\bm
H}_{RL}=\sum\limits_{\alpha=\uparrow,\downarrow}\int\limits_{RL}d^3{\bm
r} \hat\Psi^\dagger_{\alpha,RL}({\bm r})(\hat
H_{RL}-eV_{RL})\hat\Psi_{\alpha,RL}({\bm r})
 $$
are the Hamiltonians of the left and right
leads, $V_{LL,RL}$ are the electric potentials of the leads fixed
by the external voltage source,
$$
\hat {\bm
H}_{j}=\sum\limits_{\alpha=\uparrow,\downarrow}\int\limits_{j}d^3{\bm
r} \hat\Psi^\dagger_{\alpha,j}({\bm r})(\hat H_{j}-e\hat {\bm
V}_{j})\hat\Psi_{\alpha,j}({\bm r})
 $$
defines the Hamiltonians of the left ($j=L$) and
right ($j=R$) quantum dots and
$$\hat H_j=\frac{(\hat p_\mu-\frac e c A_\mu(r))^2}{2m}-\mu+U_j(r)$$
is the one-particle Hamiltonian of electron in $j$-th quantum dot
with disorder potential $U_j(r)$. Electron transfer between the
left and the right quantum dots will be described by the
Hamiltonian
$$\hat{\bm T}=\sum_{\alpha=\uparrow,\downarrow}\int_{J_1+J_2} d^2{\bm
r}\, \big[t({\bm r})\hat\Psi^\dagger_{\alpha,L}({\bm r})
\hat\Psi_{\alpha,R}({\bm r})+{\rm c.c.}\big].$$ Here the
integration runs over the total area of both tunnel barriers J$_1$
and J$_2$. The Hamiltonian $\hat {\bm T}_{L(R)}$ describing
electron transfer between the left dot and the left lead (the
right dot and the right lead) is defined analogously and are
omitted here.

Before we proceed with our analysis the following two remarks are
in order. Firstly, we point out that within our approach the
effect of electron-electron interactions is accounted for by the
voltage operators $\hat {\bm V}_{L,R}$ in the effective
Hamiltonian (\ref{H}). In order to avoid misunderstandings we
would like to emphasize that this approach is fully equivalent to
one employing the usual Coulomb interaction term in the initial
Hamiltonian of the system. The operators $\hat {\bm V}_{L,R}$
corresponding to fluctuating potentials of the left and right dots
emerge as a result of the exact Hubbard-Stratonovich decoupling of
the Coulomb term containing the product of four electron
operators. This is a standard procedure (described in details,
e.g., in Ref. \onlinecite{SZ} and elsewhere) which is bypassed
here for the sake of brevity.

Secondly, we note that in Ref. \onlinecite{GZ06} we have studied
weak localization effects in a system of coupled quantum dots
within the framework of the scattering matrix formalism combined
with the non-linear $\sigma$-model. However, in order to
incorporate interaction effects into our consideration --
similarly to Refs. \onlinecite{GZ08,GZ07} -- it will be convenient
for us to describe inter-dot electron transfer within the
tunneling Hamiltonian approach, as specified above. For clarity
let us briefly recapitulate the relation between these two
approaches. For this purpose we define the matrix elements
$t_{lm}=\langle l|\hat T| m\rangle$ between the $l-$th wave
function in the left dot and $m-$th wave function in the right
dot. Electron transfer between these dots can then be described by
a set of eigenvalues of this matrix $\tilde t_k$ where, as above,
the index $k$ labels the conducting channels. These eigenvalues
are related to the barrier channel transmissions $T_k$ as \cite{B}
\begin{equation}
T_k=\frac{4\pi^2|\tilde t_k|^2/\delta_{L}\delta_R}
{(1+\pi^2|\tilde t_k|^2/\delta_{L}\delta_R)^2}.
\end{equation}
This equation allows to keep track of the relation between two
approaches at every stage of our calculation.

We now proceed employing the path integral Keldysh technique. The
time evolution of the density matrix of our system is described by
the standard equation
\begin{equation}
\hat \rho(t)=e^{-i\hat Ht}\hat\rho_0\,e^{i\hat Ht},
\end{equation}
where $\hat H$ is given by Eq. (\ref{H}). Let us express the
operators $e^{-i\hat Ht}$ and $e^{i\hat Ht}$ via path integrals
over the fluctuating electric potentials $V_j^{F,B}$ defined
respectively on the forward and backward parts of the Keldysh
contour:
\begin{eqnarray}
e^{-i\hat Ht}&=&\int  DV_j^F\; {\rm T}\,\exp\left\{-i\int_0^t
dt'\hat H\left[V_j^F(t')\right]\right\},
\nonumber\\
e^{i\hat Ht}&=&\int  DV_j^B\; \tilde{\rm T}\,\exp\left\{i\int_0^t
dt'\hat H\left[V_j^B(t')\right]\right\}.
\end{eqnarray}
Here ${\rm T}\,\exp$ ($\tilde {\rm T}\,\exp$) stands for the time
ordered (anti-ordered) exponent and the Hamiltonians $\hat
H\left[V_j^F(t')\right]$, $\hat H\left[V_j^B(t')\right]$ are
obtained from the original Hamiltonian (\ref{H}) if one replaces
the operators ${\bf \hat V}_j(t)$ respectively by the fluctuating
voltages $V^F_j(t')$ and $V^B_j(t')$.

Let us define the effective action of our system
\begin{eqnarray}
iS[V^F,V^B]&=&\ln\left( {\rm tr} \left[ {\rm
T}\,\exp\left\{-i\int_0^t dt'\hat H\left[V_j^F(t')\right]\right\}
\right.\right. \nonumber\\ &&\times\, \left.\left. \hat\rho_0
\tilde{\rm T}\,\exp\left\{i\int_0^t dt'\hat
H\left[V_j^B(t')\right]\right\} \right]\right).
\end{eqnarray}
Since the operators $\hat H\left[V_j^F(t')\right]$, $\hat
H\left[V_j^B(t')\right]$ are quadratic in the electron creation
and annihilation operators, it is possible to integrate out the
fermionic variables and to rewrite the action in the form
\begin{equation}
    iS=iS_C+iS_{ext}+2 \bf Tr\ln \left[\check  G^{-1} \right].
\label{ac1}
\end{equation}
Here $S_C$ is the standard term describing charging effects,
$S_{ext}$ accounts for an external circuit and ${\bf\check  G^{-1}
}$ is the inverse Green-Keldysh function of electrons, moving in
fluctuating voltages field. It has the following matrix structure:
\begin{equation}
  {\bf \check G^{-1}}=\left(\begin{array}{cccc}
   \hat G^{-1}_{LL} & \hat T_L  & 0 & 0 \\
     \hat T^\dag_L & \hat G^{-1}_L & \hat T & 0 \\
     0 & \hat T^\dag & \hat G^{-1}_R & \hat T_R \\
     0 & 0 & \hat T^\dag_R & \hat G^{-1}_{RL}
\end{array}\right).
\end{equation}

Here each quantum dot as well as two leads is represented by the
2x2 matrix in the Keldysh space:
\begin{equation}
   \hat G^{-1}_i=\left(\begin{array}{cc}
    i\partial_t-\hat H_i+eV^F_i & 0 \\
     0 & -i\partial_t+\hat H_i -eV^B_i
\end{array}
\right)
\end{equation}
Tunneling blocks has the following structure in Keldysh space:
\begin{equation}
   \hat T_{L,R}=\left(\begin{array}{cc}
     -\int\limits_{J_{L,R}} t_{L,R}(r')\delta(r'-r)dr' \qquad\qquad  0\qquad\qquad \\
          \qquad\qquad  0\qquad\qquad  \int\limits_{J_{L,R}} t_{L,R}(r')\delta(r'-r)dr'
\end{array}
\right),
\end{equation}
\begin{equation}
   \hat T=\left(\begin{array}{cc}
     -\int\limits_{J_1+J_2} t(r')\delta(r'-r)dr' \qquad\qquad  0\qquad\qquad \\
          \qquad\qquad  0\qquad\qquad  \int\limits_{J_1+J_2} t(r')\delta(r'-r)dr'
\end{array}
\right).
\end{equation}

\section{Effective action}

In what follows it will be convenient for us to remove the
fluctuating voltage variables and the vector potential from the
bare Green functions. This is achieved by performing a unitary
transformation under the trace in Eq. (\ref{ac1}). As a result we
find
\begin{equation}
\hat T=\hat T_1 e^{-i\varphi_g^{(1)}}+\hat T_2
e^{-i\varphi_g^{(2)}}
\end{equation}
\vspace{0.1cm}
\begin{equation}
\hat T_l=\left(\begin{array}{c}
-e^{i\varphi_F}\int\limits_{J_l} t(r')\delta(r'-r)dr'\qquad\qquad  0\qquad\qquad \\
          \qquad\qquad0\qquad\qquad   e^{i\varphi_B}\int\limits_{J_l} t(r')\delta(r'-r)dr'
\end{array}\right).
\end{equation}
Here we introduced the fluctuating phase differences
\begin{equation}
  \varphi_{F,B}(t)=e\int^t d\tau (V^{F,B}_R(\tau)-V^{F,B}_L(\tau))
\end{equation}
defined on the forward and backward parts of the Keldysh contour
as well as the geometric phases
 \begin{equation}
  \varphi_{g}^{(1,2)}=\frac{e}{c}\int\limits_{L}^{R} dx_\mu
  A_\mu(x),
\end{equation}
where the integration contour starts in the left dot, crosses the
first ($\varphi_{g}^{(1)}$) or the second ($\varphi_{g}^{(2)}$)
junction and ends in the right dot. The difference between these
two geometric phases equals to
\begin{equation}
\varphi_g^{(1)}-\varphi_g^{(2)}=2\pi \Phi /\Phi_0,
\end{equation}
where $\Phi$ is the magnetic flux threading our system.

Let us now expand the exact action $iS$ (\ref{ac1}) in powers of
$\hat T$. Keeping the terms up to the fourth order in the
tunneling amplitude, we obtain
 \begin{eqnarray}
     iS\approx iS_C+iS_{ext}+iS_L+iS_R-2{\bf tr}\left[\hat G_L\hat T \hat G_R\hat T^\dag\right]-\nonumber\\
     -{\bf tr}\left[\hat G_L\hat T \hat G_R\hat T^\dag\hat G_L\hat T \hat G_R\hat T^\dag\right]+...
     \label{action}
 \end{eqnarray}
The terms $iS_{L,R}$ define the contributions of the isolated dots
(which are of no interest for us here), the second order terms
$\propto t^2$ yield the well known Ambegaokar-Eckern-Sch\"on (AES)
action \cite{SZ} $iS^{AES}$, and the fourth order terms $\propto
t^4$ account for the weak localization correction to the system
conductance \cite{GZ08,GZ07}.

Let us first analyze the AES action. Performing averaging of this
action over disorder in each dot separately as well as averaging
of tunneling amplitudes with the correlation function
\begin{equation}
 \overline{t({\bf x})t({\bf y})}=\frac{g_t({\bf x})}{8\pi^2 N_L N_R}\delta({\bf x}-{\bf y})
\end{equation}
we arrive at the following result
\begin{widetext}
\begin{eqnarray}
 iS^{AES}= -\int dt_1 dt_2\int\limits_{J_1+J_2} d{\bf x}\frac{g_t({\bf x})}{4\pi^2 N_L N_R}
\sum\limits_{i,j=F,B}\hat G_L^{ij}({\bf x}t_1;{\bf
x}t_2)(-1)^je^{i\varphi_j(t_2)}\hat G_R^{ji}({\bf x}t_2;{\bf
x}t_1)(-1)^ie^{-i\varphi_i(t_1)}, \label{AES}
\end{eqnarray}
where the convention $(-1)^F=-1,\ (-1)^B=1$ is implied. This AES
contribution to the action is described by the standard diagram
depicted in Fig. 2a. We observe that after disorder averaging the
AES action (\ref{AES}) becomes totally independent of the magnetic
flux. Hence, this part of the action does not account for the AB
effect investigated here.
\begin{figure}
 \centering
\includegraphics[width=3.5in]{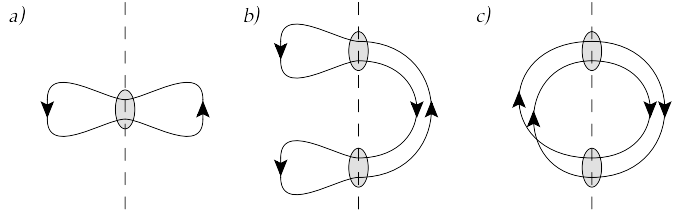}
\caption{\label{f2}Diagrammatic representation of different
contributions originating from expansion of the effective action
in powers of the central barrier transmissions: second order (AES)
terms (a) and different fourth order terms (b,c).}
\end{figure}

In order to evaluate the contribution sensitive to the magnetic
flux $\Phi$ it is necessary to analyze the last term in Eq.
(\ref{action}). Averaging over realizations of transmission
amplitudes yields two types of terms illustrated by the diagrams
in Fig. \ref{f2}b,c.  It is straightforward to check that only the
contribution generated by the diagram (c) depends on the external
magnetic flux, while the diagram (b) does not depend on $\Phi$. On
top of that, the terms originating from the diagram (b) turn out
to be parametrically small for metallic quantum dots considered
here.  This observation will be justified in Appendix A.

It follows from the above arguments that only the diagram in Fig.
2c is responsible for the AB effect in our system. Its
contribution to the action reads
\begin{eqnarray}
iS_{\Phi}=-\sum\limits_{m,n=1,2}e^{2i(\varphi_g^{(n)}-\varphi_g^{(m)})}\int
dt_1 dt_2 dt_3 dt_4 \int\limits_{J_n} d{\bf x}\int\limits_{J_m} d{\bf y}
\frac{g_t({\bf x})g_t({\bf y})}{64\pi^4 N_L^2 N_R^2}\times \nonumber\qquad\quad    \\
\times\sum\limits_{i,j,k,l=F,B}\hat G_L^{ij}({\bf x}t_1;{\bf y}t_2)(-1)^je^{i\varphi_j(t_2)}
\hat G_R^{jk}({\bf y}t_2;{\bf x}t_3)(-1)^k e^{-i\varphi_k(t_3)}\times\nonumber\\
\times \hat G_L^{kl}({\bf x}t_3;{\bf
y}t_4)(-1)^le^{i\varphi_l(t_4)}\hat G_R^{li}({\bf y}t_4;{\bf
x}t_1)(-1)^ie^{-i\varphi_i(t_1)}. \label{swl1}
\end{eqnarray}
Since $\hat G_{L,R}$ are the equilibrium Green-Keldysh functions
of the dots they can be expressed via retarded ($G^R$) and
advanced ($G^A$) Green functions in the standard manner:
\begin{equation}
\hat G_{L,R}({\bf x_1}t_1;{\bf x_2}t_2)= \int dt( G^R_{L,R}({\bf
x_1}t_1;{\bf x_2}t)\hat F_1(t-t_2)-\hat F_2(t_1-t)G^A_{L,R}({\bf
x_1}t;{\bf x_2}t_2)), \label{GRGA}
\end{equation} where
\begin{equation}
     \hat F_1(t)=\left(\begin{array}{cc}
           h(t) & -f(t) \\
           h(t) & -f(t)
     \end{array}\right),\qquad\qquad
     \hat F_2(t)=\left(\begin{array}{cc}
           -f(t) & -f(t) \\
           h(t) & h(t)
     \end{array}\right).
\end{equation}
Here $f(t)=\int f(E)dE/2\pi$ is the Fourier transform of the Fermi
function $f(E)=(\exp (E/T)+1)^{-1}$ and $h(t)=\delta(t)-f(t)$.

What remains is to combine Eqs. (\ref{GRGA}) and (\ref{swl1}) and
to average the latter over disorder. This procedure amounts to
evaluating the averages of the products of retarded and advanced
Green functions in each dot separately. Such averaging can be
conveniently accomplished either by means of the diagram technique
or with the aid of the non-linear $\sigma$-model. The
corresponding calculation is presented in Appendix A. It yields
($i=L,R$):
\begin{equation}
     \langle G_i^R({\bf x_1}t_1;{\bf x_2}t_2)G_i^R({\bf x_3}t_3;{\bf x_4}t_4)\rangle_{d}=
     \langle G_i^R({\bf x_1}t_1;{\bf x_2}t_2)\rangle_{d}\langle G_i^R({\bf x_3}t_3;{\bf x_4}t_4)\rangle_{d},
\end{equation}
\begin{equation}
     \langle G_i^A({\bf x_1}t_1;{\bf x_2}t_2)G_i^A({\bf x_3}t_3;{\bf x_4}t_4)\rangle_d=
     \langle G_i^A({\bf x_1}t_1;{\bf x_2}t_2)\rangle_d\langle G_i^A({\bf x_3}t_3;{\bf x_4}t_4)\rangle_d,
\end{equation}
\begin{eqnarray}
     \langle G_i^R({\bf x_1}t_1;{\bf x_2}t_2)G_i^A({\bf x_3}t_3;{\bf x_4}t_4)\rangle_d=
     \langle G_i^R({\bf x_1}t_1;{\bf x_2}t_2)\rangle_d\langle G_i^A({\bf x_3}t_3;{\bf x_4}t_4) \rangle_d+\nonumber\\
     +2\pi N_i \varpi (|{\bf x_1}-{\bf x_4}|)\varpi (|{\bf x_2}-{\bf x_3}|)\times\qquad\qquad\nonumber\\
     \times\mathcal D_i \left(t_1-t_2;\frac{{\bf x_1}+{\bf x_4}}{2},\frac{{\bf x_2}+{\bf x_3}}{2}\right)\delta(t_1-t_2+t_3-t_4)+\nonumber\\
     +2\pi N_i \varpi (|{\bf x_1}-{\bf x_3}|)\varpi (|{\bf x_2}-{\bf x_4}|)\times\qquad\qquad\nonumber\\
     \times\mathcal C_i \left(t_1-t_2;\frac{{\bf x_1}+{\bf x_3}}{2},\frac{{\bf x_2}+{\bf
     x_4}}{2}\right)\delta(t_1-t_2+t_3-t_4),
\label{dis_av}
\end{eqnarray}
where $\mathcal D_{L,R}(t;{\bf x},{\bf y})$ and $\mathcal
C_{L,R}(t;{\bf x},{\bf y})$ the diffusons and the Cooperons in the
left and right dots and $\varpi(r)=e^{-r/2l}\sin k_Fr/k_Fr$.
Substituting these averages into the action (\ref{swl1}) it is
straightforward to observe that only the terms containing the
product of two Cooperons yield the contribution which depends on
the magnetic flux $\Phi$. This part of the action takes the form
\begin{eqnarray}
 iS^{WL}_{\Phi}=-i\sum\limits_{m,n=1,2}e^{2i(\varphi_g^{(n)}-\varphi_g^{(m)})}\int d\tau_1 d\tau_2
 \int dt_1 dt_2 dt_3 dt_4 \int\limits_{J_n} d{\bf x}\int\limits_{J_m} d{\bf y}
 \frac{g_t({\bf x})g_t({\bf y})}{4\pi^2 N_L N_R}\times \nonumber\\
\times\mathcal C_L(\tau_1;{\bf y},{\bf x})\mathcal C_R(\tau_2;{\bf
x},{\bf y})
e^{i(\varphi^+(t_2)-\varphi^+(t_3)+\varphi^+(t_4)-\varphi^+(t_1))}\sin\frac{\varphi^-(t_1)}{2}\times\nonumber
\\ \times
\left[h(t_1-t_2-\tau_1)e^{i\frac{\varphi^-(t_2)}{2}}+f(t_1-t_2-\tau_1)e^{-i\frac{\varphi^-(t_2)}{2}}\right]\times\nonumber\\
\times
\left[h(t_2-t_3-\tau_2)e^{-i\frac{\varphi^-(t_3)}{2}}f(t_3-t_4+\tau_1)-\right.
\quad\qquad\qquad\nonumber\\ \left.-f(t_2-t_3-\tau_2)e^{i\frac{\varphi^-(t_3)}{2}}h(t_3-t_4+\tau_1)    \right]\times\nonumber\\
\times \left[e^{i\frac{\varphi^-(t_4)}{2}}f(t_4-t_1+\tau_2)+e^{-i\frac{\varphi^-(t_4)}{2}}h(t_4-t_1+\tau_2)\right]+\nonumber\\
+\{L\leftrightarrow R,\varphi^{\pm}\rightarrow -\varphi^{\pm}\},
\label{sw2}
\end{eqnarray}
\end{widetext}
where we defined the ``classical'' and the ``quantum'' components
of the fluctuating phase:
\begin{equation}
   \varphi^+(t)=\frac{\varphi_F(t)+\varphi_B(t)}{2},\qquad\varphi^-(t)=\varphi_F(t)-\varphi_B(t).
\end{equation}
The above expression for the action $S^{WL}_{\Phi}$ (\ref{sw2})
fully accounts for coherent oscillations of the system conductance
in the lowest non-vanishing order in tunneling. It is important to
emphasize that no additional approximations were employed during
its derivation and, in particular, the fluctuating phases are {\it
exactly} accounted for. We will make use of this fact in the next
section while considering the effect of electron-electron
interactions on AB oscillations in the system under consideration.

\section{Current oscillations}
Let us now evaluate the current $I$ through our system. For this
purpose we will employ a general formula
\begin{equation}
  I=-e\int\mathcal D^2\varphi^{\pm} \frac{\delta S[\varphi^+,\varphi^-]}{\delta \varphi^-(t)} e^{iS[\varphi^+,\varphi^-]}.
\end{equation}
Substituting the total effective action into this formula we
arrive at the result for the current which can be split into two
terms $I=I_0+\delta I$, where $I_0$ is the flux-independent
contribution and $\delta I$ is the quantum correction to the
current sensitive to the magnetic flux $\Phi$. This correction is
determined by the action $iS^{WL}_{\Phi}$, i.e.
\begin{equation}
 \delta I=-e\int\mathcal D^2\varphi^{\pm} \frac{\delta S^{WL}_{\Phi}[\varphi^+,\varphi^-]}{\delta \varphi^-(t)}
 e^{iS[\varphi^+,\varphi^-]}.
\label{IAB}
\end{equation}
Below we will only be interested in finding the quantum correction
(\ref{IAB}).

In order to evaluate the path integral over the phases
$\varphi^{\pm}$ in (\ref{IAB}) we note that the contributions
$S_C$ and $S_{\rm ext}$ in Eq. (\ref{action}) are quadratic in the
fluctuating phases provided our external circuit consists of
linear elements. Other contributions to the action are, strictly
speaking, non-Gaussian. However, in the interesting for us here
metallic limit (\ref{met}) phase fluctuations can be considered
small down to exponentially low energies \cite{PZ91,Naz99} in
which case it suffices to expand both contributions  up to the
second order $\varphi^{\pm}$. Moreover, this Gaussian
approximation becomes {\it exact}  \cite{GGZ05} in the limit of
fully open left and right barriers with $g_{L,R} \gg 1$. Thus, in
the metallic limit (\ref{met}) the integral (\ref{IAB}) remains
Gaussian at all relevant energies and can easily be performed.

This task can be accomplished with the aid of the following
correlation functions
\begin{equation}
   \langle\varphi^+(t)\rangle=eVt,\qquad
   \langle\varphi^-(t)\rangle=0,
\label{cf1}
\end{equation}
\begin{equation}
   \langle (\varphi^+(t)-\varphi^+(0))\varphi^+(0)\rangle=-F(t),
\label{cf2}
\end{equation}
\begin{equation}
   \langle \varphi^+(t)\varphi^-(0)+\varphi^-(t)\varphi^+(0)\rangle =2iK(|t|),
\label{cf3}
\end{equation}
\begin{equation}
   \langle \varphi^+(t)\varphi^-(0)-\varphi^-(t)\varphi^+(0)\rangle=2iK(t),
\label{cf4}
\end{equation}
\begin{equation}
   \langle \varphi^-(t)\varphi^-(0)\rangle=0,
\label{cf5}
\end{equation}
where the last relation follows directly from the causality
principle \cite{GZ98}. Here and below we define $V=V_{RL}-V_{LL}$ to
be the transport voltage across our system.

Substituting the AB action (\ref{sw2}) into Eq. (\ref{IAB}) one
arrives at the expression containing six different phase averages
listed in Appendix B.  All these averages in Eqs.
(\ref{B1})-(\ref{B6}) are expressed in terms of two real
correlation functions $F(t)=
\langle(\hat\varphi(t)-\hat\varphi(0))^2\rangle /2$ and
$K(t)=i\langle[\hat\varphi(0),\hat\varphi(t)]\rangle /2$ defined
above in Eqs. (\ref{cf2}) and (\ref{cf3}). Note that these
correlation functions are well familiar from the so-called
$P(E)$-theory\cite{SZ,IN} describing electron tunneling in the
presence of an external environment which can also mimic
electron-electron interactions in metallic conductors. They are
expressed in terms of an effective impedance $Z(\omega)$ ``seen''
by the central barriers J$_1$ and J$_2$
\begin{equation}
   F(t)=e^2\int\frac{d\omega}{2\pi}\coth\frac{\omega}{2T}\Re[Z(\omega)]\frac{1-\cos(\omega
   t)}{\omega},
\label{Ft}
\end{equation}
\begin{equation}
  K(t)=e^2\int\frac{d\omega}{2\pi}\Re[Z(\omega)]\frac{\sin(\omega
  t)}{\omega}.
\label{Kt}
\end{equation}
Further evaluation of these correlation functions for our system
is straightforward and yields
\begin{equation}
F(t)\simeq \frac{4}{g} \left(\ln\left|\frac{\sinh(\pi T t)}{\pi
 T\tau_{RC}}\right|+\gamma    \right),
\label{FFF}
\end{equation}
\begin{equation}
K(t)\simeq\frac{2\pi}{g}{\rm sign}(t), \label{KKK}
\end{equation}
where we defined $g=4\pi/e^2Z(0)$ and $\gamma\simeq0.577$ is the
Euler constant. Neglecting the contribution of external leads
(which can be trivially restored if needed) and making use of the
inequality (\ref{met}) we obtain $g\simeq 2g_Lg_R/(g_L+g_R)$.

We observe that while $F(t)$ grows with time at any temperature
including $T=0$, the function $K(t)$ always remains small in the
limit $g\gg 1$ considered here. As we demonstrate in Appendix B,
the correlation function $F(t)$ should be fully kept in the
exponent in Eqs. (\ref{B1})-(\ref{B6}) while the correlator $K(t)$
can be safely ignored in the leading order in $1/g$. Then
combining all terms we observe that the Fermi function $f(E)$ -- though
present in the effective action (\ref{sw2}) -- drops out from the
final expression for the quantum correction to the current which
takes the form:
\begin{widetext}
\begin{eqnarray}
   \delta I (\Phi )=-\sum\limits_{m,n=L,R}\frac{e^2 Ve^{2i(\varphi_g^{(n)}-\varphi_g^{(m)})}}{8\pi^3N_L N_R}\int d\tau_1 d\tau_2
\int\limits_{J_n} d{\bf x}\int\limits_{J_m} d{\bf y}g_t({\bf x})g_t({\bf y})\times\nonumber\\
\times\mathcal C_L(\tau_1;{\bf y},{\bf x})\mathcal C_R(\tau_2;{\bf
x},{\bf
y})e^{-2F(\tau_1)-2F(\tau_2)+F(\tau_1-\tau_2)+F(\tau_1+\tau_2)}.
\label{IABee}
\end{eqnarray}
\end{widetext}
We observe that the amplitude of AB oscillations is affected by
the electron-electron interaction only via the correlation
functions for the ``classical'' component of the
Hubbard-Stratonovich phase $\varphi^+$. Both the correlators
containing the ``quantum'' phase $\varphi^-$ and the Fermi
function $f(E)$ enter only in the next order in $1/g$ which
defines weak Coulomb correction to $\delta I$ ignored here. For
more details on this point we refer the reader to Refs.
\onlinecite{GZ08,GZ07}.

The result (\ref{IABee}) can also be rewritten as
\begin{equation}
  \delta I(\Phi )=-I_{AB}(\Phi )-I_{WL1}-I_{WL2},
\label{AB+q}
\end{equation}
where the first -- flux dependent -- term in the right-hand side
explicitly accounts for AB oscillations and reads
\begin{equation}
I_{AB}(\Phi )=I_{AB}\cos(4\pi\Phi/\Phi_0),
\label{ABAB}
\end{equation}
while the last two terms $I_{WL1,2}$ represent the remaining part
of the quantum correction to the current which does not depend on
$\Phi$.

Already at this stage we would like to clarify the relation
between our present results for AB oscillations and those for WL
correction to conductance \cite{GZ08}. In order to derive Eq.
(\ref{IABee}) we have evaluated the contributions of all processes
illustrated by the diagrams in Fig. 2b,c and identified terms
sensitive to the magnetic field which were not considered in Ref.
\onlinecite{GZ08}.  In this way we have obtained the AB current
$I_{AB}(\Phi)$ in Eqs. (\ref{AB+q}), (\ref{ABAB}) which represents
our new result to be analyzed below. The two remaining terms in
Eq. (\ref{AB+q}) are the WL corrections already evaluated in Ref.
\onlinecite{GZ08}. Towards the end of this section we will
explicitly specify the relation between all three contributions to
the quantum correction (\ref{AB+q}).

Let us evaluate the amplitude of AB oscillations $I_{AB}$ for the
system with two identical quantum dots with volume $\mathcal V$,
dwell time $\tau_{D}$ and dimensionless conductances $g_L=g_R
\equiv g=4\pi/\delta\tau_D$, where $\delta=1/ \mathcal V N$ is the
dot mean level spacing and $N$ is the electron density of states.
In this case the Cooperons take the form
\begin{equation}
 \mathcal C_L(t;{\bf x},{\bf y})=\mathcal C_R(t;{\bf x},{\bf y})=\frac{\theta(t)}{\mathcal
 V}e^{-t/\tau_D}.
\label{coop}
\end{equation}
Defining dimensionless conductances of central barriers as
$g_{t1,2}=\int_{J_{1,2}}g_t(x)dx$ we obtain
\begin{widetext}
\begin{eqnarray}
   I_{AB}=\frac{e^2g_{t1}g_{t2}\delta^2V}{4\pi^3}\int\limits_0^\infty d\tau_1 d\tau_2
   e^{-\frac{\tau_1+\tau_2}{\tau_D}-2F(\tau_1)-2F(\tau_2)+F(\tau_1-\tau_2)+F(\tau_1+\tau_2)}.
   \label{res}
\end{eqnarray}
\end{widetext}
In the absence of electron-electron interactions ($F(\tau ) \to
0$) this formula yields:
\begin{equation}
I_{AB}^{(0)}=\frac{4e^2 g_{t1} g_{t2}V}{\pi g^2}. \label{nonint}
\end{equation}

\begin{figure}[t]
 \centering
\includegraphics[width=3.5in]{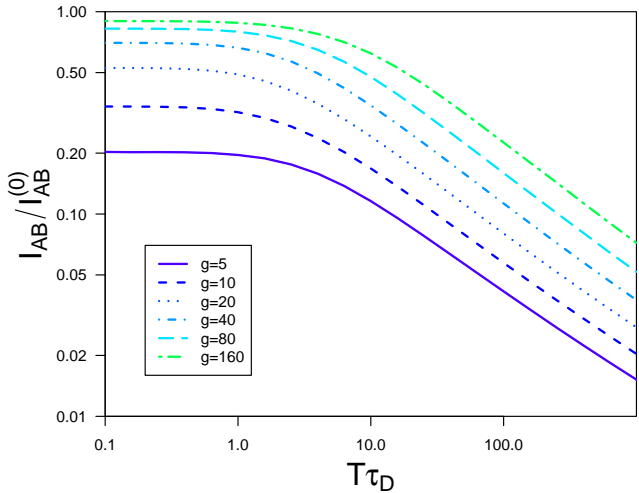}
\caption{\label{f3} (Color online) The ratio $I_{AB}/I_{AB}^{(0)}$
versus temperature at different values of dimensionless
conductance $g$.}
\end{figure}

In order to account for the effect of interactions we need to
specify the effective impedance $Z(\omega )$. Its real part takes
the form
\begin{equation}
\Re Z(\omega)= \frac{4\pi}{e^2
g}\left(\frac{\tau^2}{\tau^2_{RC}}\frac{1}{\omega^2\tau^2+1}+\frac{\pi\delta(\omega)}{\tau_D+\tau_{RC}}\right),
\label{rez}
\end{equation}
where $1/\tau=1/\tau_D+1/\tau_{RC}$, $\tau_{RC}=\pi/gE_C$ is the
$RC$-time and $E_C$ is an effective charging energy of our system.
Eq. (\ref{res}) demonstrates that electron-electron interactions
always tend to suppress the amplitude of AB oscillations $I_{AB}$
below its non-interacting value (\ref{nonint}). Combining Eqs.
(\ref{FFF}) and (\ref{rez}) with (\ref{res}) at high enough
temperatures we obtain
\begin{equation}
 \frac{ I_{AB}}{I_{AB}^{(0)}}=\left\{\begin{array}{lc}
  e^{-\frac{8\gamma}{g}}\frac{(2\pi T\tau_{RC})^{8/g}}{1+4\pi T\tau_D/g},\quad &    \tau_D^{-1} \lesssim T \lesssim \tau_{RC}^{-1}, \\
 \frac{1}{2\tau_D}\left(\frac{g\tau_{RC}}{T}\right)^{1/2}, &  \tau_{RC}^{-1} \lesssim
 T,
  \end{array}\right.
\end{equation}
while in the low temperature limit we find
\begin{equation}
 \frac{I_{AB}}{I_{AB}^{(0)}}=e^{-\frac{8\gamma}{g}}\left(\frac{2\tau_{RC}}{\tau_D}\right)^{8/g}, \qquad T\lesssim
 \tau_D^{-1}.
 \end{equation}
The latter result demonstrates that interaction-induced
suppression of AB oscillations in metallic dots with $\tau_{RC}
\ll \tau_D$ persists down to $T=0$.

 \begin{figure}[t]
 \centering
\includegraphics[width=3.5in]{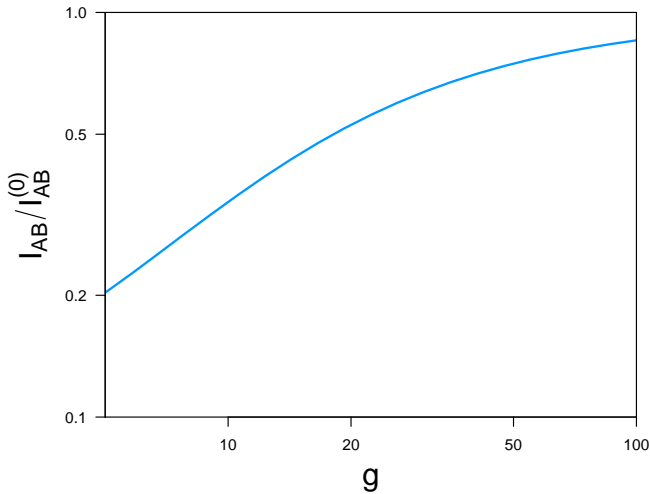}
\caption{\label{f4}(Color online) The ratio $I_{AB}/I_{AB}^{(0)}$
as a function of dimensionless conductance $g$ at $T=0$ and
$\tau_{D}/\tau_{RC}=10$.}
\end{figure}

The ratio $I_{AB}/I_{AB}^{(0)}$ was also evaluated numerically as
a function of temperature at different values of $g$. The
corresponding results are presented in Fig. \ref{f3}. We observe
that -- in accordance with the above analytic expressions -- the
ratio $I_{AB}/I_{AB}^{(0)}$ grows with decreasing $T$ as a power
law and finally saturates to a constant value smaller than unity
at $T \lesssim 1/\tau_D$. The suppression of AB oscillations --
both at higher temperatures and at $T \to 0$ clearly depends on
the interaction strength which is controlled by the parameter
$1/g$ in our model. Fig. \ref{f4} demonstrates the dependence of
$I_{AB}/I_{AB}^{(0)}$ on $g$ in the limit of zero temperature and
for $\tau_{D}/\tau_{RC}=10$. While at moderate values of $g \sim
10 \div 20$ interaction-induced suppression of $I_{AB}$ remains
pronounced down to $T=0$, at weaker interactions ($g \gtrsim 100$)
this effect becomes less significant and is merely important at
higher temperatures, cf. Fig. \ref{f3}.

In order to complete our analysis let us briefly address
additional quantum corrections to the current $I_{WL1,2}$ in Eq.
(\ref{AB+q}). Although these terms do not depend on $\Phi$ and,
hence, are irrelevant for AB oscillations, they allow to
establish a direct and transparent relation between the
Aharonov-Bohm effect studied here and the phenomenon of weak
localization in systems of metallic quantum dots with electron-electron
interactions \cite{GZ08,GZ07}. With the aid of Eq. (\ref{IABee})
one easily finds
\begin{equation}
\frac{I_{WL1}}{I_{AB}}=\frac{g_{t1}}{2g_{t2}}, \;\;\;\;
\frac{I_{WL2}}{I_{AB}}=\frac{g_{t2}}{2g_{t1}}. \label{WL12}
\end{equation}
Combining this equation with the above results for $I_{AB}$ we
immediately identify the terms $I_{WL1}$ and $I_{WL2}$ as weak
localization corrections to the current \cite{GZ08,GZ07}
originating from the two central barriers in our structure. In
addition, in the absence of the magnetic field $\Phi =0$ the total
quantum correction to the current $\delta I(0)$ (\ref{AB+q})
exactly coincides with the weak localization correction to the
current for two connected in series metallic quantum dots
\cite{GZ08,GZ07} provided the two central barriers in Fig. 1 are
viewed as a composite tunnel barrier with total dimensionless
conductance $g_{t1}+g_{t2}$.

\section{Concluding remarks}

The established relation between our present results and those
obtained in Refs. \onlinecite{GZ08,GZ07} helps to clarify the main
physical reason for the effect of interaction-induced suppression
of AB oscillations in our structure. In full analogy with the weak
localization correction \cite{GZ08,GZ07} both at non-zero
temperatures and $T=0$ this suppression is due to electron
dephasing by electron-electron interactions. This decoherence
effect reduces the electron ability to interfere and, hence,
decreases the amplitude $I_{AB}$ below its non-interacting value
$I_{AB}^{(0)}$. At the same time Coulomb blockade effect --
although yields an additional suppression of $I_{AB}$  -- remains
weak in metallic quantum dots and can be neglected as compared to
the dominating effect of electron dephasing. It is also important
to emphasize that in the course of our analysis we employed {\it
only one} significant approximation: We performed a regular
expansion of the current in powers of the tunneling conductances
up to second order terms (forth order terms in the tunneling
matrix elements). At the same time the effect of electron-electron
interactions on AB oscillations in our system was treated
non-perturbatively to {\it all} orders and essentially {\it
exactly}.

Note that one could be tempted to interpret the suppression of
$I_{AB}$ at $T=0$ just as a result of a simple renormalization
effect by electron-electron interactions which is not related to
dephasing. It is important to stress that -- unlike, e.g., in the
case of the interaction correction for single quantum dots
\cite{GZ01,BN} -- here such interpretation would not be
appropriate. The fundamental reason is that the interaction of an
electron with an effective environment (produced by other
electrons) effectively breaks down the time-reversal symmetry and,
hence, causes both dissipation and dephasing for interacting
electrons down to $T=0$ \cite{GZ98}. In this respect it is also
important to point out a deep relation between interaction-induced
electron decoherence and the $P(E)$-theory \cite{SZ,IN} which we
already emphasized elsewhere \cite{GZ99,GZ08,GZ07} and which is
also evident from our present results. Similarly to
\cite{GZ08,GZ07} one can also introduce the electron dephasing
time in our problem and demonstrate that at $T \to 0$ it saturates
to a finite value in agreement with available experimental
observations \cite{Pivin,Huibers,Hackens}. We believe that the
quantum dot rings considered here can be directly used for further
experimental investigations of quantum coherence of interacting
electrons in nanoscale conductors at low temperatures. We also
note that our model can possibly be applied to analyze the
behavior of recently fabricated self-assembled quantum rings
\cite{Bel} where the AB oscillations have been observed by means
of magnetization experiments.

\vspace{0.3cm}

\centerline{\bf Acknowledgments}

\vspace{0.3cm}

One of us (A.G.S.) acknowledges support from the Landau Foundation and from
the Dynasty Foundation.

\appendix
\begin{widetext}
\section{Averaging over disorder}
Let us consider the following disorder averages of the product for
retarded and advanced Green functions for one of the quantum dots:
\begin{equation}
X_d({\bf x_1},{\bf x_2};\varepsilon)= \langle G^R({\bf x_1},{\bf
x_2};\omega)G^A({\bf x_2},{\bf
x_1};\omega-\varepsilon)\rangle_d-\langle G^R({\bf x_1},{\bf
x_2};\omega)\rangle_d\langle G^A({\bf x_2},{\bf
x_1};\omega-\varepsilon)\rangle_d, \label{A1}
\end{equation}
\begin{equation}
X_c({\bf x_1},{\bf x_2};\varepsilon)= \langle G^R({\bf x_1},{\bf
x_2};\omega)G^A({\bf x_1},{\bf
x_2};\omega-\varepsilon)\rangle_d-\langle G^R({\bf x_1},{\bf
x_2};\omega)\rangle_d\langle G^A({\bf x_1},{\bf
x_2};\omega-\varepsilon)\rangle_d, \label{A2}
\end{equation}
\begin{equation}
X_k({\bf x_1},{\bf x_2};\varepsilon)= \langle G^R({\bf x_1},{\bf
x_1};\omega)G^A({\bf x_2},{\bf
x_2};\omega-\varepsilon)\rangle_d-\langle G^R({\bf x_1},{\bf
x_1};\omega)\rangle_d\langle G^A({\bf x_2},{\bf
x_2};\omega-\varepsilon)\rangle_d, \label{A3}
\end{equation}
where
\begin{equation}
  G^{R(A)}({\bf x_1}t_1;{\bf x_2}t_2)=\int\frac{d\omega}{2\pi}e^{-i\omega(t_1-t_2)}G^{R(A)}({\bf x_1},{\bf x_2};\omega).
\label{A4}
\end{equation}

\begin{figure}[t]
\parbox{5.5in}{
\includegraphics[width=1.5in,angle=270]{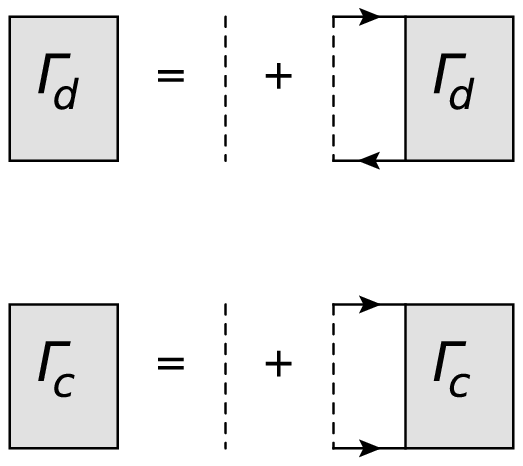}\qquad\qquad
\includegraphics[width=1.5in,angle=270]{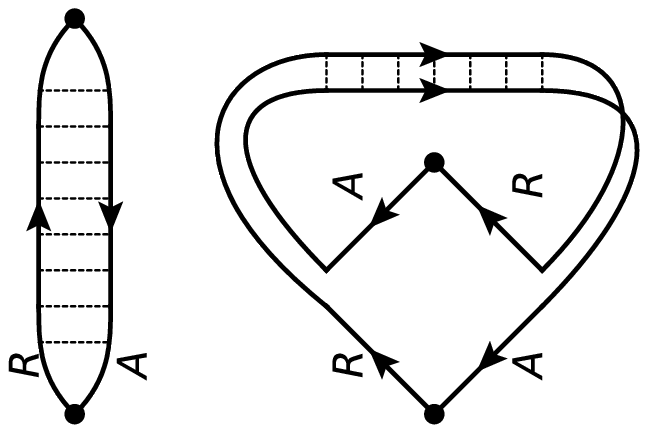}
\includegraphics[width=1.5in,angle=270]{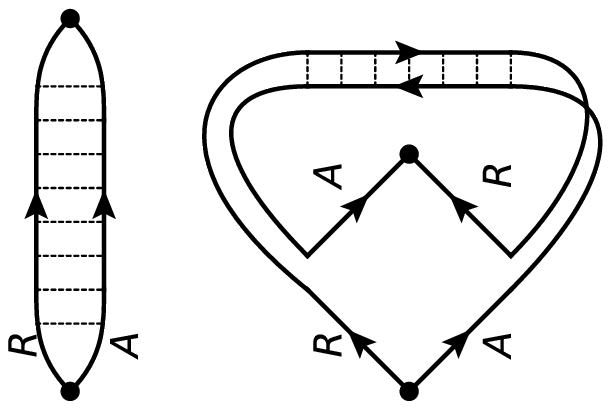}}
\caption{\label{f5}Diagrammatic representation for vertices
$\Gamma_d({\bf x_1},{\bf x_2};\omega)$, $\Gamma_c({\bf x_1},{\bf
x_2};\omega)$ and for averages $X_d({\bf x_1},{\bf
x_2};\varepsilon)$ and $X_c({\bf x_1},{\bf x_2};\varepsilon)$.}
\end{figure}

In order to evaluate the above averages we will employ the
standard diagram technique for noninteracting electrons in
disordered systems \cite{AGD}. The essential elements here are the
so-called diffuson and Cooperon ladders depicted in Fig. \ref{f5}
where we also define vertices $\Gamma_d({\bf x_1},{\bf
x_2};\omega)$ and $\Gamma_c({\bf x_1},{\bf x_2};\omega)$. In the
presence of time-reversal symmetry and in the limit of low momenta
and frequencies these vertices obey a diffusion-like equation:
\begin{equation}
(-i\omega-D\nabla^2_{{\bf x_2}})\Gamma_{d(c)}({\bf x_1},{\bf
x_2};\omega)=\frac{1}{2\pi N\tau_e^2}\delta({\bf x_1-x_2}).
\label{A5}
\end{equation}
Here $D=v_Fl/3$ and $\tau_e=l/v_F$ are respectively the diffusion
coefficient and the electron elastic mean free time. With the aid
of the above vertices one can define the diffuson and the Cooperon
respectively as
\begin{equation}
\mathcal D(t;{\bf x_1},{\bf x_2})=2\pi
N\tau_e^2\int\frac{d\omega}{2\pi}e^{-i\omega t}\Gamma_d({\bf
x_1},{\bf x_2};\omega), \label{A6}
\end{equation}
\begin{equation}
  \mathcal C(t;{\bf x_1},{\bf x_2})=2\pi N\tau_e^2\int\frac{d\omega}{2\pi}e^{-i\omega t}\Gamma_c({\bf x_1},{\bf
  x_2};\omega).
\label{A7}
\end{equation}
In the absence of the magnetic field they
obey the following diffusion equations
\begin{equation}
(\partial_t-D\nabla^2_{{\bf x_2}})\mathcal D(t;{\bf x_1},{\bf
x_2})=\delta({\bf x_1-x_2})\delta(t),
\label{A8}
\end{equation}
\begin{equation}
(\partial_t-D\nabla^2_{{\bf x_2}})\mathcal C(t;{\bf x_1},{\bf
x_2})=\delta({\bf x_1-x_2})\delta(t)
 \label{A9}
 \end{equation}
 with appropriate boundary conditions.

Evaluating the diagrams for $X_d({\bf x_1},{\bf x_2};\varepsilon)$
depicted in Fig. \ref{f5} after some algebra we arrive at the
following result:
\begin{equation}
  X_d({\bf x_1},{\bf x_2};\varepsilon)=(2\pi N\tau_e)^2\Gamma_d({\bf x_1},{\bf x_2};\varepsilon)
  +(2\pi N\tau_e)^2\varpi^2(|{\bf x_1}-{\bf x_2}|)\Gamma_c\left(\frac{{\bf x_1+x_2}}{2},\frac{{\bf
  x_1+x_2}}{2};\varepsilon\right),
\label{A10}
\end{equation}
where
\begin{equation}
\varpi(|{\bf x_1}-{\bf x_2}|)=\frac{1}{2\pi N \tau_e}\int d{\bf x}
\langle G^R({\bf x_1},{\bf x};\omega)\rangle_d\langle G^A({\bf
x},{\bf x_2};\omega)\rangle_d. \label{A11}
\end{equation}
In the case of 3d systems we find  $\varpi(r)=e^{-r/2l}\sin
k_Fr/k_Fr$.

The expression for $X_c({\bf x_1},{\bf x_2};\varepsilon)$ is
derived analogously. We find
\begin{equation}
  X_c({\bf x_1},{\bf x_2};\varepsilon)=
  (2\pi N\tau_e)^2\Gamma_c({\bf x_1},{\bf x_2};\varepsilon)+(2\pi N\tau_e)^2\varpi^2(|{\bf x_1}-
  {\bf x_2}|)\Gamma_d\left(\frac{{\bf x_1+x_2}}{2},\frac{{\bf
  x_1+x_2}}{2};\varepsilon\right).
\label{A12}
\end{equation} Combining Eqs. (\ref{A1}), (\ref{A2},
(\ref{A4}), (\ref{A6}), (\ref{A7}) with (\ref{A10})-(\ref{A12}) we
arrive at Eq. (\ref{dis_av}).

\begin{figure}[t]
\parbox{5.5in}{
\includegraphics[width=1.8in,angle=270]{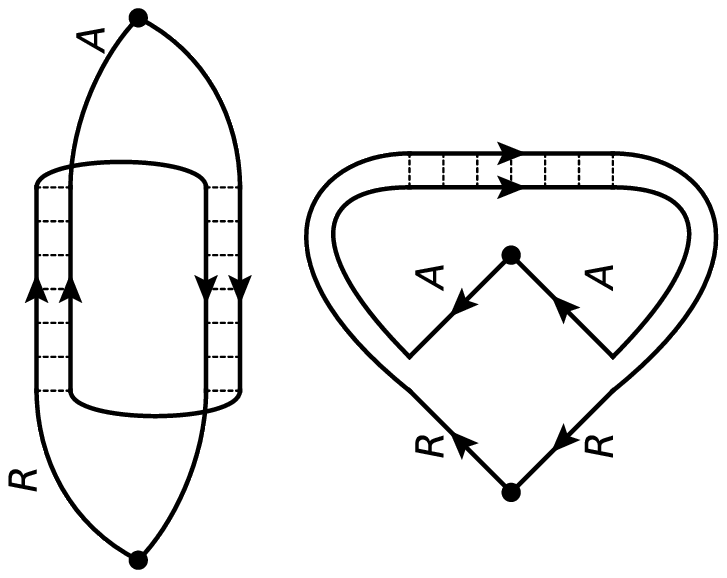}\qquad\qquad
\includegraphics[width=1.8in,angle=270]{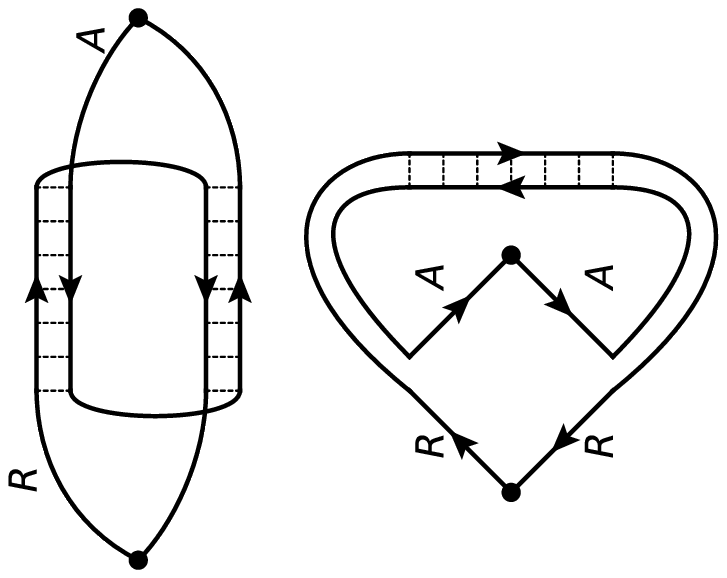}}
\caption{\label{f6}Diagrams which define the average $X_k({\bf
x_1},{\bf x_2};\varepsilon)$.}
\end{figure}

Note that the average $X_k({\bf x_1},{\bf x_2};\varepsilon)$
(\ref{A3}) is omitted in Eq. (\ref{dis_av}) since this average
turns out to be parametrically small as compared to both
$X_{d}({\bf x_1},{\bf x_2};\varepsilon)$ and $X_{c}({\bf x_1},{\bf
x_2};\varepsilon)$. In order to demonstrate this fact it is
necessary to evaluate the diagrams for $X_k({\bf x_1},{\bf
x_2};\varepsilon)$ depicted in Fig. \ref{f6}. Proceeding as above
we get
\begin{eqnarray}
X_k({\bf x_1},{\bf x_2};\varepsilon)=(2\pi
N\tau_e)^2\varpi^2(|{\bf x_1}-{\bf
x_2}|)\left[\Gamma_d\left(\frac{{\bf x_1+x_2}}{2},\frac{{\bf
x_1+x_2}}{2};\varepsilon\right)+\Gamma_c\left(\frac{{\bf
x_1+x_2}}{2},\frac{{\bf
x_1+x_2}}{2};\varepsilon\right)\right]+\nonumber\\+(2\pi
N\tau_e^2)^2\left[\Gamma_d^2({\bf x_1},{\bf
x_2};\varepsilon)+\Gamma_c^2({\bf x_1},{\bf
x_2};\varepsilon)\right]. \label{A13}
\end{eqnarray}
The first term in this equation clearly vanishes for $|{\bf
x_1}-{\bf x_2}|\gtrsim l$. Here we assume that the size of both
the dots and the contacts is large as compared to the electron
mean free path $l$. Provided the typical contact size is of the
same order as that of the dots the latter condition implies $l \ll
v_F\tau_D$. If, however, the contact size is much smaller than
that of the dots this condition becomes $p_Fl \ll \sqrt{N_{ch}}$,
where $N_{ch}$ is the effective number of conducting channels in
the contact. In this case both the diffuson and the Cooperon do
not depend on coordinates and are defined by Eq. (\ref{coop}).
Then one finds
$$\Gamma_{d(c)}(\omega)=(2\pi N\tau_e\mathcal
V(-i\omega+1/\tau_D))^{-1}.$$ Substituting this result into Eq.
(\ref{A13}) we get $X_k(t)\approx \mathcal
V^{-2}\theta(t)te^{-t/\tau_D}$. Comparing this expression with
that for $X_d(c)(t)\approx 2\pi N\mathcal
V^{-1}\theta(t)e^{-t/\tau_D}$ at times $t \lesssim \tau_D$ we
obtain
\begin{equation}
X_k/X_{d(c)}\lesssim\tau_D/(N\mathcal V)\sim \tau_D\delta \sim 1/g
\ll 1. \label{A14}
\end{equation}
This estimate demonstrates that the average $X_k$ (\ref{A3}) can
be safely disregarded in Eq. (\ref{dis_av}) for the problem under
consideration.

\section{Averaging over fluctuating phases}
Substituting the action (\ref{sw2}) into Eq. (\ref{IAB}) one
expresses the current $\delta I (\Phi )$ as a combination of
different phase averages evaluated with the total action
$S[\varphi^+,\varphi^-]$. As we already argued above, in the
metallic limit (\ref{met}) all these averages are essentially
Gaussian and, hence, can be easily performed. For the sake of
completeness, below we present the corresponding results:
\begin{eqnarray}
 \langle e^{i(\varphi^+(t_2)-\varphi^+(t_3)+\varphi^+(t_4)-\varphi^+(t_1)
+\frac{\varphi^-(t_1)}{2}+\frac{\varphi^-(t_2)}{2}+\frac{\varphi^-(t_3)}{2}+\frac{\varphi^-(t_4)}{2})
} \rangle=\nonumber\\
e^{-F(t_1-t_2)-F(t_1-t_4)-F(t_2-t_3)-F(t_3-t_4)+F(t_1-t_3)+F(t_2-t_4)}\times\nonumber\\
\times
e^{-iK(t_1-t_2)-iK(|t_1-t_3|)-iK(t_1-t_4)+iK(t_2-t_3)+iK(|t_2-t_4|)-iK(t_3-t_4)},
\label{B1}
\end{eqnarray}
\begin{eqnarray}
 \langle e^{i(\varphi^+(t_2)-\varphi^+(t_3)+\varphi^+(t_4)-\varphi^+(t_1)
+\frac{\varphi^-(t_1)}{2}+\frac{\varphi^-(t_2)}{2}+\frac{\varphi^-(t_3)}{2}-\frac{\varphi^-(t_4)}{2})
} \rangle=\nonumber\\
e^{-F(t_1-t_2)-F(t_1-t_4)-F(t_2-t_3)-F(t_3-t_4)+F(t_1-t_3)+F(t_2-t_4)}\times\nonumber\\
\times
e^{-iK(t_1-t_2)-iK(|t_1-t_3|)-iK(|t_1-t_4|)+iK(t_2-t_3)-iK(t_2-t_4)+iK(|t_3-t_4|)},
\end{eqnarray}
\begin{eqnarray}
 \langle e^{i(\varphi^+(t_2)-\varphi^+(t_3)+\varphi^+(t_4)-\varphi^+(t_1)
+\frac{\varphi^-(t_1)}{2}+\frac{\varphi^-(t_2)}{2}-\frac{\varphi^-(t_3)}{2}+\frac{\varphi^-(t_4)}{2})
} \rangle=\nonumber\\
e^{-F(t_1-t_2)-F(t_1-t_4)-F(t_2-t_3)-F(t_3-t_4)+F(t_1-t_3)+F(t_2-t_4)}\times\nonumber\\
\times
e^{-iK(t_1-t_2)+iK(t_1-t_3)-iK(t_1-t_4)-iK(|t_2-t_3|)+iK(|t_2-t_4|)-iK(|t_3-t_4|)},
\end{eqnarray}
\begin{eqnarray}
 \langle e^{i(\varphi^+(t_2)-\varphi^+(t_3)+\varphi^+(t_4)-\varphi^+(t_1)
+\frac{+\varphi^-(t_1)}{2}-\frac{\varphi^-(t_2)}{2}+\frac{\varphi^-(t_3)}{2}-\frac{\varphi^-(t_4)}{2})
} \rangle=\nonumber\\
e^{-F(t_1-t_2)-F(t_1-t_4)-F(t_2-t_3)-F(t_3-t_4)+F(t_1-t_3)+F(t_2-t_4)}\times\nonumber\\
\times
e^{-iK(t_1-t_2)-iK(|t_1-t_3|)+iK(|t_1-t_4|)+iK(|t_2-t_3|)-iK(|t_2-t_4|)+iK(|t_3-t_4|)},
\end{eqnarray}
\begin{eqnarray}
 \langle e^{i(\varphi^+(t_2)-\varphi^+(t_3)+\varphi^+(t_4)-\varphi^+(t_1)
+\frac{-\varphi^-(t_1)}{2}+\frac{\varphi^-(t_2)}{2}-\frac{\varphi^-(t_3)}{2}+\frac{\varphi^-(t_4)}{2})
} \rangle=\nonumber\\
e^{-F(t_1-t_2)-F(t_1-t_4)-F(t_2-t_3)-F(t_3-t_4)+F(t_1-t_3)+F(t_2-t_4)}\times\nonumber\\
\times e^{-iK(|t
_1-t_2|)+iK(|t_1-t_3|)-iK(|t_1-t_4|)-iK(|t_2-t_3|)+iK(|t_2-t_4|)-iK(|t_3-t_4|)},
\end{eqnarray}
\begin{eqnarray}
 \langle e^{i(\varphi^+(t_2)-\varphi^+(t_3)+\varphi^+(t_4)-\varphi^+(t_1)
+\frac{\varphi^-(t_1)}{2}+\frac{\varphi^-(t_2)}{2}-\frac{\varphi^-(t_3)}{2}-\frac{\varphi^-(t_4)}{2})
} \rangle=\nonumber\\
e^{-F(t_1-t_2)-F(t_1-t_4)-F(t_2-t_3)-F(t_3-t_4)+F(t_1-t_3)+F(t_2-t_4)}\times\nonumber\\
\times
e^{-iK(t_1-t_2)+iK(t_1-t_3)+iK(|t_1-t_4|)+iK(|t_2-t_3|)-iK(|t_2-t_4|)+iK(|t_3-t_4|)}.
\label{B6}
\end{eqnarray}
\end{widetext}
Note that for arbitrary metallic conductors $g_{L,R} \gg 1$ all
these equations are accurate down to exponentially small energies
(set by the so-called renormalized charging energy
\cite{PZ91,Naz99} which is of little importance for us here), and
in the particular limit of fully open left and right barriers Eqs.
(\ref{B1})-(\ref{B6}) become exact \cite{GGZ05}. Thus, combining
(\ref{B1})-(\ref{B6}) with Eqs. (\ref{sw2}), (\ref{IAB}),
(\ref{FFF}) and (\ref{KKK}) we exactly account for the effect of
electron-electron interactions on the amplitude of AB oscillations
in the system under consideration.

It is useful to observe that in order to quantitatively describe
this effect in the metallic limit $g \gg 1$ one can totally
neglect all the functions $K(t)$ in all Eqs.
(\ref{B1})-(\ref{B6}). This is because these functions remain much
smaller than one at all times (cf. Eq. (\ref{KKK})) and, hence,
can only cause a weak ($\sim 1/g$) Coulomb correction to
$I_{AB}$ which further slightly decreases the amplitude of AB
oscillations. The origin of this Coulomb correction is exactly the
same as that identified and discussed in the weak localization
problem \cite{GZ08,GZ07}. Thus, no additional discussion of this
point is necessary here.

Substituting unity instead of all the exponents in
(\ref{B1})-(\ref{B6}) containing $K$-functions and keeping all
$F$-functions in the exponent, one easily arrives at Eq. (\ref{IABee}).


\begin{thebibliography}{30}
\bibitem{ArSh} For a review see, e.g., A.G. Aronov and Yu.V. Sharvin,
Rev. Mod. Phys. {\bf 59}, 755 (1987).
\bibitem{CS} For a review see, e.g., S. Chakravarty and A. Schmid,
Phys. Rep. {\bf 140}, 193 (1986).
\bibitem{GZ06} D.S. Golubev and A.D. Zaikin, Phys. Rev. B {\bf 74}, 245329 (2006).
\bibitem{GZ08} D.S. Golubev and A.D. Zaikin, New J. Phys. {\bf 10}, 063027 (2008).
\bibitem{GZ07} D.S. Golubev and A.D. Zaikin, Physica E {\bf 40}, 32 (2007).
\bibitem{SZ} G. Sch\"on and A.D. Zaikin, Phys. Rep. {\bf 198}, 237 (1990).
\bibitem{IN} G.L. Ingold and Yu.V. Nazarov, {\it Single Charge Tunneling},
 (Plenum Press, New York) {\it NATO ASI Series} B {\bf 294}, p. 21 (1992).
\bibitem{GZ99} D.S. Golubev and A.D. Zaikin,  {\it Quantum Physics at
Mesoscopic Scale} (EDP Sciences, Les Ulis, 2000), p. 491.
\bibitem{B} C.W.J. Beenakker, Rev. Mod. Phys. {\bf 69}, 731 (1997).
\bibitem{PZ91}  S.V. Panyukov and A.D. Zaikin, Phys. Rev. Lett. {\bf 67}, 3168 (1991).
\bibitem{Naz99} Yu.V. Nazarov,  Phys. Rev. Lett. {\bf 82}, 1245 (1999).
\bibitem{GGZ05} D.S. Golubev, A.V. Galaktionov, and A.D. Zaikin, Phys. Rev. B {\bf
    72}, 205417 (2005).
\bibitem{GZ98} D.S. Golubev and A.D. Zaikin, Phys. Rev. Lett. {\bf 81}, 1074 (1998);
Phys. Rev. B {\bf 59}, 9195 (1999); {\it ibid.} {\bf 62}, 14061
(2000); Physica B {\bf 255}, 164 (1998).
\bibitem{GZ01} D.S. Golubev and A.D. Zaikin, Phys. Rev. Lett. {\bf 86}, 4887 (2001);
Phys. Rev. B {\bf 69}, 075318 (2004).
\bibitem{BN} D.A. Bagrets and Yu.V. Nazarov, Phys. Rev. Lett. {\bf 94}, 056801 (2005).
\bibitem{Pivin} D.P. Pivin, A. Andresen, J.P. Bird, and D.K. Ferry, Phys. Rev. Lett. {\bf 82}, 4687 (1999).
\bibitem{Huibers} A.G. Huibers, J.A. Folk, S.R. Patel, C.M. Marcus, C.I. Duruoz, and J.S. Harris,
Phys. Rev. Lett. {\bf 83}, 5090 (1999).
\bibitem{Hackens} B. Hackens, S. Faniel, C. Gustin, X. Wallart, S. Bollaert,
A. Cappy, and V. Bayot, Phys. Rev. Lett. {\bf 94}, 146802 (2005).
\bibitem{Bel} N.A.J.M. Kleemans, I.M.A. Bominaar-Silkens, V.M.
Fomin, V.N. Gladilin, D. Granados, A.G. Taboada, J.M. Garcia, P.
Offermans, U. Zeitler, P.C.M. Christianen, J.C. Maan, J.T.
Devreese, and P.M. Koenraad, Phys. Rev. Lett. {\bf 99}, 146808
(2007).
\bibitem{AGD}  A.A. Abrikosov, L.P. Gorkov, and I.Ye. Dzyaloshinski, {\it
    Quantum Field Theoretical Methods in Statistical Physics}, 2nd ed.
(Pergamon, Oxford, 1965).
\end{thebibliography}
\end{document}